\documentclass[prl,aps,nofootinbib,twocolumn,preprintnumbers,10pt]{revtex4}
\usepackage{amsmath,amssymb,graphicx,epsfig,hyperref,breakurl,color}
\usepackage{lipsum}

\begin{document}

\title{ 
Novel method to reliably  determine the photon helicity in $b\to s\gamma$ }

\author{ Wei Wang$^1$~\footnote{Corresponding author, Email:wei.wang@sjtu.edu.cn},
Fu-Sheng Yu$^2$~\footnote{Corresponding author, Email:yufsh@lzu.edu.cn} and Zhen-Xing
Zhao~$^{1,3}$\footnote{Corresponding author, Email:zhaozx15@163.com} }

\affiliation{
$^1$ INPAC, SKLPPC, MOE KLPPC,
School of Physics and Astronomy, Shanghai Jiao Tong University, Shanghai 200240, China \\
$^2$School of Nuclear Science and Technology, Lanzhou University, Lanzhou 730000,   China \\
$^3$School of Physical Science and Technology, Inner Mongolia University, Hohhot 010021, China }

\begin{abstract}
A sizable  right-handed photon  polarization in $b\to s\gamma$ is a clear signal for new physics. 
In this work we point out that the photon helicity in $b\to s\gamma$ can be unambiguously   extracted  by combining   the measurements in $B\to K_1\gamma$ and the Cabibbo-favored  $D\to K_1 e^+\nu$ decay. We propose a ratio of up-down asymmetries in   $D\to K_1 e^+\nu$ to quantify the hadronic effects.    { A method for measuring   in experiment the involved  partial decay widths in the ratio is discussed, and experimental facilities like BESIII,  Belle-II and LHCb are likely to  measure this ratio. We also give the angular distribution that is useful to extract the photon polarization in the presence of different kaon resonances. }
\end{abstract}


\pacs{xxxx}

\maketitle

\textit{Introduction.}\textemdash{}
Nowadays searching for new physics is the primary objective in particle physics. 
In the standard model (SM), 
the photon helicity in   $b\to s\gamma$ decay is  predominantly left-handed and thereby its measurement  plays a unique role in probing right-handed couplings in new physics (NP)~\cite{Atwood:1997zr,Becirevic:2012dx,Paul:2016urs}. A representative  example is the left-right symmetric model~\cite{Kou:2013gna,Haba:2015gwa},  in which the photon can acquire a significant right-handed component. However to date there are not many   experimental  results on the photon helicity.

It is noticed that the photon helicity is related to an up-down asymmetry $\mathcal{A}_{\rm UD}$ in $B\to K_1 \gamma$~\cite{Gronau:2001ng,Gronau:2002rz,Kou:2010kn}~\footnote{Throughout this paper the $K_1$ denotes the axial-vector meson $K_1(1270)$ and $K_1(1410)$. } and more generally the angular distribution in $B\to K_{res}(\to K\pi\pi)\gamma$.  However   measuring the up-down asymmetry    in $B\to K_1 \gamma$~\cite{Aaij:2014wgo} does  not directly reveal  the photon  helicity  since  the detailed knowledge of $K_1 \to  K\pi\pi$ is a prerequisite. Previous  theoretical analyses have adopted nonperturbative models to parametrize the $K_1\to K\pi\pi$ amplitude and thus  considerable hadronic uncertainties are inevitably introduced~\cite{Gronau:2001ng,Gronau:2002rz,Kou:2010kn,Tayduganov:2011ui,Gronau:2017kyq}.  

In this work we will point out  that one can combine the measurements in $B\to K_1\gamma$ and semileptonic $D\to  K_1 l^+\nu(l=e,\mu)$ decays to determine  the photon polarization in $b\to s\gamma$ without any theoretical ambiguity.  In particular, we propose a ratio of up-down asymmetries in   $D\to K_1 e^+\nu$, ${\cal A}_{UD}^{\prime}$,  to quantify the hadronic effects in $K_1\to K\pi\pi$ decay and   point out that the photon helicity can be expressed in terms of $\mathcal{A}_{\rm UD}$  and ${\cal A}_{UD}^{\prime}$.  Experimental facilities including  BESIII,  Belle-II and LHCb are likely to  measure this ratio ${\cal A}_{UD}^{\prime}$.


\textit{Photon polarization in $B\to K_{1}(\to K\pi\pi)\gamma$.}\textemdash Let
us start with the angular distribution in $B\to K_{1}(\to K\pi\pi)\gamma$. The effective Hamiltonian for $b\to s\gamma$ has the general form: 
\begin{align}
{\cal H}_{{\rm eff}} & =-\frac{4G_{F}}{\sqrt{2}}V_{tb}V_{ts}^{*}(C_{7L}{\cal O}_{7L}+C_{7R}{\cal O}_{7R}),\nonumber \\
 & \qquad{\cal O}_{7L,R}=\frac{em_{b}}{16\pi^{2}}\bar{s}\sigma_{\mu\nu}\frac{1\pm\gamma_{5}}{2}bF^{\mu\nu},
\end{align}
where $C_{7L,7R}$   are the Wilson coefficients  for ${\cal O}_{7L,R}$.
Due to the chiral structure of $W^{\pm}$ couplings to quarks in the
SM,  the emitted photon in $b\to s\gamma$ is mostly left-handed   and the right-handed configuration is suppressed by $C_{7R}^{\rm SM}/C_{7L}^{\rm SM}\approx m_{s}/m_{b}$. For $\bar b\to \bar s\gamma$, it is vice versa.

The   differential decay rate for $ {B}\to K_{1}(\to {K}\pi\pi)\gamma$
can be expressed  as a sum of contributions from left- and right-polarized
photons~\cite{Gronau:2001ng,Gronau:2002rz,Gronau:2017kyq}: 
\begin{eqnarray}\label{eq:BK1g}
&&\frac{d\Gamma_{ K_1\gamma}}{d\cos\theta_K}=\frac{|A|^2|\vec{J}|^{2}}{4}  \nonumber\\
&&\times  \left[1+\cos^{2}\theta_K+2\lambda_{\gamma}\cos\theta_K\frac{{\rm Im}[\vec{n}\cdot(\vec{J}\times\vec{J}^{*})]}{|\vec{J}|^{2}}\right]. 
\end{eqnarray}
Hereafter the $\theta_K$ is chosen as the relative angle between the normal direction $\vec{n}$ of  the $K_1$ decay plane and the opposite flight direction of photon in $K_1$ rest frame.  The coefficient $A$ is a nonperturbative amplitude. 
The $\vec J$ characterizes the $K_1\to K\pi\pi$ decay amplitude with ${\cal A} (K_{1}\to K\pi\pi) = \vec \epsilon_{K_{1}}\cdot \vec J$. The $\cos\theta_K$ is a parity-odd quantity, and the left-handed and right-handed polarizations  contribute with an opposite  sign. 
The    parameter $\lambda_{\gamma}$ is defined as
\begin{equation}\label{eq:lambdagamma}
\lambda_{\gamma}\equiv\frac{|\mathcal{A}(  B\to   K_{1R}\gamma_R)|^{2}-|\mathcal{A}(  B\to   K_{1L}\gamma_L)|^{2}}{|\mathcal{A}(  B\to   K_{1R}\gamma_R)|^{2}+|\mathcal{A}(  B\to   K_{1L}\gamma_L)|^{2}}, 
\end{equation}
with  $\lambda_{\gamma}\simeq-1$ for $b\to s\gamma$ and $\lambda_{\gamma}\simeq+1$ for $\bar b\to \bar s\gamma$    in the SM.  

Compared to the angular distribution in the above equation, an integrated up-down asymmetry  is more convenient on the experimental side~\cite{Gronau:2001ng}: 
\begin{align}\label{eq:updown-K1gamma}
{\cal A}_{\rm UD} & \equiv\frac{ \Gamma_{K_1\gamma}[\cos\theta_K>0]-\Gamma_{K_1\gamma}[\cos\theta_K<0]}{ \Gamma_{K_1\gamma}[\cos\theta_K>0]+\Gamma_{K_1\gamma}[\cos\theta_K<0]}  \\
 & =\lambda_{\gamma}\frac{3}{4}\frac{{\rm Im}[\vec{n}\cdot(\vec{J}\times\vec{J}^{*})]}{|\vec{J}|^{2}}. \nonumber
\end{align} 
The LHCb collaboration has  measured the up-down asymmetry in $B^+\to K^+\pi^-\pi^+\gamma$~\cite{Aaij:2014wgo}  with $\mathcal{A}_{\rm UD}=(6.9\pm1.7)\times 10^{-2}$ in the range of $m_{K\pi\pi}=[1.1, 1.3]$GeV.   { In this kinematics region it is expected that the asymmetry is dominated by $K_1(1270)$ but other contributions might also be important. } Nevertheless 
it can be seen that,  { even assuming the dominance of $K_1(1270)$}  it is also essential  to fathom
the hadronic factor ${\rm Im}[\vec{n}\cdot(\vec{J}\times\vec{J}^{*})]/|\vec{J}|^{2}$. 
Many estimations on this input factor have been made in either model-dependent or phenomenological approaches~\cite{Gronau:2001ng,Gronau:2002rz,Kou:2010kn,Tayduganov:2011ui,Gronau:2017kyq,Kou:2016iau}.
Unfortunately, the current understanding of $K_1\to K\pi\pi$ is very limited, due to the complicated intermediate states of $K^*\pi$, $K\rho$, $(K\pi)_{\rm S-wave}\pi$ and $K(\pi\pi)_{\rm S-wave}$, and their phases for interferences. Considerable hadronic  uncertainties are thus inevitably introduced and beyond control. Therefore, the accurate result of $\lambda_\gamma$ has never been achieved so far, even though the up-down asymmetry has been well measured. 


\textit{Determination of photon helicity by combining $B\to K_1\gamma$ and    $D\to K_1e^+ \nu_e$ decays.}\textemdash  We now proceed with the angular distribution for $D\to K_{1}(\to {K}\pi\pi)e^+\nu$, and  demonstrate that  combining the measurements in $B\to K_1\gamma$ and    $D\to K_1e^+ \nu_e$ can determine the photon helicity in $b\to s\gamma$  in a model-independent way.

\begin{figure}\begin{center}
\includegraphics[scale=0.5]{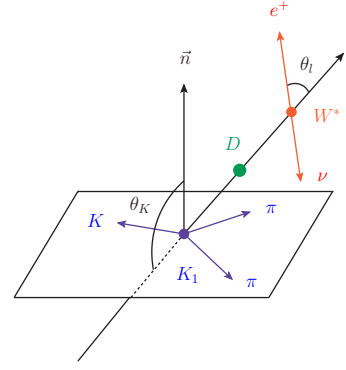}
\caption{ Kinematics for $D\to K_{res}(\to K\pi\pi) e^+\nu$. The relative angle between the normal direction of $K_{res}$ decay plane and the opposite  of $D$ flight direction in the $K_{res}$ rest frame is denoted as $\theta_K$, while  the $\theta_l$ is introduced  as  the relative angle between the flight directions of $e^{+}$ in the $e^+\nu$ rest frame and the $e^+\nu$ in the $D$ rest frame.  } \label{fig:kinematics}
\end{center}
\end{figure}

With the kinematics shown in  Fig.~\ref{fig:kinematics}.
one can derive  the angular distribution for $D\to K_1(\to K\pi\pi ) e^+ \nu_e$ as: 
\begin{align}
\frac{d\Gamma_{ K_1e\nu_e}}{d\cos\theta_K d\cos\theta_l} &=  d_{1}[1+\cos^{2}\theta_K\cos^{2}\theta_{l}]\nonumber \\
 & \; +d_{2}[1+\cos^{2}\theta_K]\cos\theta_{l}\nonumber \\
 &\; +d_{3}\cos\theta_K[1+\cos^{2}\theta_{l}]\nonumber \\
 &\; +d_{4}\cos\theta_K\cos\theta_{l}\nonumber \\
 &\;  +d_{5}[\cos^{2}\theta_K+\cos^{2}\theta_{l}].  \label{eq:angularK1e+nu}
\end{align}The angular coefficients are given as: 
\begin{align}
d_{1} & =\frac{1}{2}|\vec{J}|^{2}(4|c_{0}|^{2}+|c_{-}|^{2}+|c_{+}|^{2}),\nonumber \\
d_{2}   &=-|\vec{J}|^{2}(|c_{-}|^{2}-|c_{+}|^{2}),\nonumber \\
d_{3} & =-{\rm Im}\,[\vec{n}\cdot(\vec{J}\times\vec{J}^{*})](|c_{-}|^{2}-|c_{+}|^{2}),\nonumber \\
d_{4} & =2{\rm Im}\,[\vec{n}\cdot(\vec{J}\times\vec{J}^{*})](|c_{-}|^{2}+|c_{+}|^{2}),\nonumber \\
d_{5} & =-\frac{1}{2}|\vec{J}|^{2}(4|c_{0}|^{2}-|c_{-}|^{2}-|c_{+}|^{2}). 
\end{align} 
In the above   we have  neglected the lepton mass  and the $c_{0,+,-}$ corresponds to the nonperturbative amplitudes for $D$ decays into  $K_1$ with different polarizations.   

Compared to the angular distribution for $B\to K_1\gamma$, the one for $D\to K_1e^+\nu$ is different in three aspects. In $B\to K_1\gamma$, the emitted  photon is onshell and thus only transversely polarizations are allowed, but the longitudinal polarization also exists in $D\to K_1e^+\nu$, contributing with the amplitude $c_0$. Secondly   while only one angle $\theta_K$ is constructed for $B\to K_1\gamma$, two angles  $\theta_K$ and $\theta_l$ are involved  in $D\to K_1e^+\nu$. Thirdly,   in $B\to K_1\gamma$, the  $\cos\theta_K$ itself is a parity-odd quantity. In Eq.~\eqref{eq:BK1g},  the parity-even term $(1+\cos^2\theta_K)$ is accompanied with $|\vec J|^2$, and the parity-odd one contains the decay factor ${\rm Im}[\vec{n}\cdot(\vec{J}\times\vec{J}^{*})]$.  
In $D\to K_1e^+\nu$,  the $\cos\theta_K$ dependence is similar, but  the lepton pair is produced  through the $V-A$ current. This interaction also gives the parity-even term,  and parity-odd term in $\cos\theta_l$. We have picked up the two parity-odd terms, formed by the $\cos\theta_K(1+\cos^2\theta_l)$  that is proportional to ${\rm Im}[\vec{n}\cdot(\vec{J}\times\vec{J}^{*})]$ and the $\cos\theta_l(1+\cos^2\theta_K)$ that is proportional to $|\vec{J}|^{2}$.  {\it The ratio of the coefficients of these two terms, namely $d_3$ and $d_2$,  give the requested hadronic factor ${\rm Im}[\vec{n}\cdot(\vec{J}\times\vec{J}^{*})]/|\vec{J}|^{2}$.}

To pick up the $d_2$ and $d_3$ in a simpler way,  we  also propose to explore  a ratio  of  up-down asymmetries (or forward-backward asymmetries): 
\begin{align}\label{eq:AUD2}
{\cal A}_{\rm UD}^{\prime}  & \equiv \equiv\frac{ \Gamma_{K_1e\nu_e}[\cos\theta_K>0]-\Gamma_{K_1e\nu_e}[\cos\theta_K<0]}{ \Gamma_{K_1e\nu_e}[\cos\theta_l>0]-\Gamma_{K_1e\nu_e}[\cos\theta_l<0]}. 
\end{align}
It is straightforward to find   
\begin{align}
{\cal A}_{\rm UD}^{\prime}  &=\frac{{\rm Im}[\vec{n}\cdot(\vec{J}\times\vec{J}^{*})]}{|\vec{J}|^{2}}. \label{eq:updown-Klnu}
\end{align} 
Apparently  quantifying the ${\cal A}_{\rm UD}$ in $B\to K_1\gamma$  and  ${\cal A}_{\rm UD}^{\prime} $ in $D\to K_1e^+\nu$  in experiment  will  help to extract  the photon helicity in $b\to s\gamma$:
\begin{eqnarray}
\lambda_{\gamma}= \frac{4}{3} \frac{{\cal A}_{UD}}{{\cal A}_{\rm UD}^{\prime} }.  \label{eq:combinationpolarization}
\end{eqnarray}


The $D\to K_1(1270)e^+\nu$ channel is a Cabibbo-favored decay process and thus its decay branching fraction is expected sizable. 
On the experimental side, an earlier evidence for $D^0\to K^-_1(1270)e^+\nu$ has been found by CLEO in Ref.~\cite{Artuso:2007wa}. 
Quite recently, using the $2.93{\rm fb}^{-1}$  data sample   of $e^+e^-$ collision at the center of mass energy of $3.773$ GeV, BESIII collaboration has observed  $D^+\to \overline K^0_1(1270)e^+\nu$ for the first time  with a statistical significance greater than $10\sigma$ and  the measured branching fraction is \cite{Ablikim:2019wxs}:
\begin{align}
{\cal B}(D^+\to \overline K^0_1 e^+\nu) = (2.3\pm0.26^{+0.18}_{-0.21}\pm0.25)\times 10^{-3}. 
\end{align}
From this measured branching fraction, one may infer that  a direct measurement of the ${\cal A}_{UD}^{\prime}$ is feasible with more statistics  in the near future.  We also expect that a high precision could be achieved taking into account the fact that  much more data  will  be accumulated at BESIII, Belle-II, and LHCb, leaving aside the Super tau-charm factory in the future. 

 {\textit{Experimental implementation. }\textemdash{} The analysis in the previous section is focused on only the  $K_1$ contribution.   
Though the LHCb measurement of the spectrum of $K\pi\pi$ in  $B\to K\pi\pi\gamma$~\cite{Aaij:2014wgo} and an earlier measurement by Belle~\cite{Yang:2004as}  indicate  that the $K_1(1270)$ contributions dominates in the range of $m_{K\pi\pi}=[1.1, 1.3]$GeV, other contributions are likely non-negligible. In the  next section, we will derive  the angular distribution with different kaon resonances, and here we  will briefly  discuss the implementation  in experiment. }

 { As shown in  Eqs.~\eqref{eq:updown-K1gamma} and \eqref{eq:AUD2},   the up-down asymmetries are  constructed using   decay widths into $K_1$ with $\cos\theta_K>0$ and $\cos\theta_K<0$.  In experiment it is possible to  divide  the $B\to K\pi\pi \gamma $  and  $D\to K\pi\pi e^+\nu  $ decay widths   into two regions, with $\cos\theta_K>0$ and $\cos\theta_K<0$ respectively.  In both regions, one can make a fit of the $K\pi\pi$ spectrum by including different kaon resonances  and then the involved  decay widths  for $K_1$ with  $\cos\theta_K>0$ and $\cos\theta_K<0$ can be  obtained.   An analysis  of  the total decay width of $B\to K\pi\pi\gamma$ has been  conducted by Belle collaboration~\cite{Yang:2004as}, deriving the branching ratio ${\cal B}(B^+\to K_1^+(1270)\gamma)= (4.3\pm0.9\pm0.9)\times 10^{-5}$.    Based on the large amount of data accumulated by LHCb collaboration and the future Belle-II experiment,  it is expected that the involved  decay widths for decays into $K_1$ with $\cos\theta_K>0$ and $\cos\theta_K<0$ can be obtained by such an analysis.  With these partial widths, the up-down asymmetries  can be obtained and accordingly the photon polarization can be used to probe/constrain new physics models.   }


\textit{Photon polarization in $B\to K\pi\pi\gamma$ and $D\to K\pi\pi e^+\nu$. }\textemdash{}  It is also meaningful to include contributions from more $K_J$ resonances, and  in particular  the $K_1(1400),K_2(1430)$ contributions will interfere with that from the $K_1(1270)$.    A vector $K^*(1410)$ resonance will not contribute to the photon helicity measurement, and thus its contribution is not shown in the following.

With the $K_1,K_2$ resonating contributions,  the $B\to K\pi\pi\gamma$ angular distribution now becomes: 
\begin{eqnarray}\label{eq:BtoKpipigamma}
& & \frac{d\Gamma(B\to K\pi\pi\gamma)}{d\cos\theta_K} =  \frac{d\Gamma_{K_1\gamma}}{d\cos\theta_K}   \nonumber\\
&&+ \frac14  |B|^2 |\vec K\,|^2 \big\{
 (\cos^2\theta_K + \cos^2 2\theta_K )  \nonumber\\
&& + 
\lambda_\gamma\frac{2 \mbox{Im}[\vec n\cdot (\vec K\times\vec K\,^*)]}{ |\vec K\,|^2}
\cos\theta_K\cos 2\theta_K\big\} \nonumber \\
&&+ \mbox{Im}[AB^* \vec n\cdot (\vec 
J\times \vec K\,^*)] \big\{\frac12 (3\cos^2\theta_K-1)  \nonumber\\
&& 
+ \lambda_\gamma \frac{\mbox{Re}[AB^*(\vec J\cdot \vec K^*)]}{\mbox{Im}[AB^* \vec n\cdot (\vec 
J\times \vec K\,^*)]} \cos^3\theta_K
\big\}.
\end{eqnarray}
The  $(B, \vec K)$  are  nonperturbative coefficients relating to $K_2(1430)$, and their explicit forms can be found in Ref.~\cite{Gronau:2002rz}.  As shown in the above, the photon polarization $\lambda_{\gamma}$ can also be extracted through the $K_2$ contribution  in the third line of above equation   or the $K_1-K_2$ interference term in the fifth line. But again such determinations request the knowledge of nonperbative matrix elements,   $\mbox{Im}[\vec n\cdot (\vec K\times\vec K\,^*)]/ |\vec K\,|^2$, and  $\mbox{Re}[AB^*(\vec J\cdot \vec K^*)]/\mbox{Im}[AB^* \vec n\cdot (\vec 
J\times \vec K\,^*)]$. 

Including the resonances, we also give the angular  distributions for $D\to K_{res}(\to K\pi\pi) e^+\nu$: 
\begin{align}
& & \frac{d\Gamma(D\to K\pi\pi e\nu_e)}{d\cos\theta_Kd\cos\theta_l} =\sum_{K_J= K_1,  K_2,K_{12}^I}
 \frac{d\Gamma_{ K_Je\nu_e}}{d\cos\theta_Kd\cos\theta_l}. 
 \end{align} 
The $D\to K_2(\to K\pi\pi)e^+\nu$ contribution is:  
\begin{align}
&&  \frac{d\Gamma_{K_2e\nu_e}}{d\cos\theta_K d\cos\theta_l} =   |c_0'|^2 \frac{3}{2} \sin^2(2\theta_K) \sin^2\theta_l |\vec K|^2 \nonumber\\
&& + 2 |c_+'|^2  |\vec K\,|^2  \cos^4\frac{\theta_l}{2}  \big\{
 (\cos^2\theta_{K} + \cos^2 2\theta_{K} )  \nonumber\\
  && + 
2 \cos\theta_{K}\cos 2\theta_{K} \frac{\mbox{Im}[\vec n\cdot (\vec K\times\vec K\,^*)]}{  |\vec K\,|^2} 
\big\} \nonumber\\
&& +2  |c_{-}'|^2   |\vec K\,|^2 \sin^4\frac{\theta_l}{2}  \big\{
(\cos^2\theta_{K} + \cos^2 2\theta_{K} ) \nonumber\\
&&   -
2 \cos\theta_{K}\cos 2\theta_{K} \frac{\mbox{Im}[\vec n\cdot (\vec K\times\vec K\,^*)]}{  |\vec K\,|^2} 
\big\},  
\end{align} 
The coefficients in the third and fifth line mimic the requested input $\mbox{Im}[\vec n\cdot (\vec K\times\vec K\,^*)]/ |\vec K\,|^2$ requested in the third line of Eq.~\eqref{eq:BtoKpipigamma}. 
The $K_1-K_2$ interference is given as:
\begin{eqnarray}
 &  & \frac{d {\Gamma}_{ K_{12}^Ie\nu_e}}{d\cos\theta_{K}d\cos\theta_{l}} \nonumber\\&& =-4\sqrt{3}\sin^{2}(\theta_{K})\cos\theta_{K}\sin^{2}\theta_{l}{\rm Re}[c_{0}(c_{0}')^{*}\vec{J}\cdot\vec{K}^{*}]\nonumber \\
 &  & -8\cos^{4}\frac{\theta_{l}}{2} \mbox{Im}[c_{+}(c_{+}')^{*}\vec{n}\cdot(\vec{J}\times\vec{K}\,^{*})] \big\{ \frac{1}{2}(3\cos^{2}\theta_K-1)\nonumber \\
 &  &+\cos^{3}\theta_K\frac{\mbox{Re}[c_{+}(c_{+}')^{*} (\vec{J}\cdot\vec{K}^{*})]}{\mbox{Im}[c_{+}(c_{+}')^{*}\vec{n}\cdot(\vec{J}\times\vec{K}\,^{*})]} \big\} \nonumber \\
 &  & -8\sin^{4}\frac{\theta_{l}}{2}\mbox{Im}[c_{-}(c_{-}')^{*}\vec{n}\cdot(\vec{J}\times\vec{K}\,^{*})]\big\{  \frac{1}{2}(1-3\cos^{2}\theta_K)\nonumber \\
 &  &+\cos^{3}\theta_K\frac{\mbox{Re}[c_{+}(c_{+}')^{*} (\vec{J}\cdot\vec{K}^{*})]}{\mbox{Im}[c_{+}(c_{+}')^{*}\vec{n}\cdot(\vec{J}\times\vec{K}\,^{*})]} \big\}. 
\end{eqnarray}
The $c^{\prime}_{0,+,-}$ correspond to the nonperturbative amplitudes for $D$ decays into the  $K_2$.  {In this interference term the relation between $\frac{\mbox{Re}[c_{+}(c_{+}')^{*} (\vec{J}\cdot\vec{K}^{*})]}{\mbox{Im}[c_{+}(c_{+}')^{*}\vec{n}\cdot(\vec{J}\times\vec{K}\,^{*})]} $  and $\propto AB^*$ is less obvious.      }

\textit{Discussions.}\textemdash{}   
 Although the lepton mass has been neglected, we have checked that our method through the angular distribution analysis  is still valid when the lepton is massive. Thus this analysis also applies to  $D\to K_{res}(\to K\pi\pi)\mu^+\nu_\mu$. 

In the above, we have elucidated the method using the angular distribution of $D\to K_{res}(\to K\pi\pi)l^+\nu$ decay, but one can also use the $B_s\to K_{res}(\to K\pi\pi)l\bar\nu$,  $D_s\to K_{res}(\to K\pi\pi)l^+\nu$  decays and $\tau\to K_{res}(\to K\pi\pi)\nu$. An estimate the branching fraction of  $B_s\to K_1l\bar\nu$  is about $(3.65^{+2.27}_{-1.87})\times 10^{-4}$~\cite{Li:2009tx}, and  might be measured in the future. The $D_s\to K_{res} l^+\nu$ is a $c\to d$ transition suppressed by the CKM matrix element, and needs more data.

Combining the measurements in $B\to K\pi\pi\gamma$ and $D\to K\pi\pi\ell\nu$ can give  the absolute value of $|C_{7R}/C_{7L}|$ via $\lambda_\gamma$ in Eq. (\ref{eq:lambdagamma}).
This  constraint on $C_{7R}/C_{7L}$ is complementary to those from the time-dependent $CP$ asymmetries in  $B^0\to f_{CP}\gamma$ (where $f_{CP}$ is the $CP$-eigenstate)~\cite{Atwood:1997zr,Atwood:2004jj,Akar:2018zhv} which measure $S_{f_{CP}\gamma}\propto\mathcal{I}m[e^{-i\phi}C_{7R}/C_{7L}]$ (where $\phi$ is the phase in the $B^0-\overline B^0$ mixing), and the angular distributions in $B \to K^* (\to K\pi)\gamma(\to e^+e^-)$ \cite{Grossman:2000rk,Kruger:2005ep,Becirevic:2011bp} with $A_T^{(2)}(0)\propto\mathcal{R}e[C_{7R}/C_{7L}]$ and $A_T^{(im)}(0)\propto\mathcal{I}m[C_{7R}/C_{7L}]$. 

Photon helicity and the right-handed couplings are similar for  $b\to s\gamma$ and $b\to d\gamma$ in the SM, but might be different in NP models. Thus   the $B\to a_1(1260)\gamma\to  \pi\pi\pi\gamma$  is also of great interest, and  combining the measurements  $B\to a_1\gamma$ and $D\to a_1e^+\nu$ will allow to determine  the photon helicity in $b\to d\gamma$  in a model-independent way.

\textit{Conclusions.}\textemdash{}  
Since photons emitted  in  $b\to s\gamma$
decay are predominantly  left-handed,  measuring photon polarizations in this mode can   test the  standard model and probe the new physics effects with large right-handed couplings. 
It has been noticed that the photon polarization in $b\to s\gamma$ is related to the up-down asymmetry in  $B\to K_1\gamma$. But unfortunately  this observation  does not provide a full understanding of photon  polarization, since  this  requests  the knowledge on  the  $K_{1}\to K\pi\pi$ decay, and introduces the uncontrollable  model-dependence.

In this work we have  pointed  out  that one can combine the measurements in $B\to K_1\gamma$ and semileptonic $D\to  K_1 l^+\nu(l=e,\mu)$ decays and determine  the photon polarization in $b\to s\gamma$ without any theoretical ambiguity.  In particular, we proposed a ratio of up-down asymmetries in   $D\to K_1 e^+\nu$, ${\cal A}_{UD}^{\prime}$,  to quantify the hadronic effects in $K_1\to K\pi\pi$ decay.  Experimental facilities including  BESIII,  Belle-II and LHCb are likely to  measure this ratio in the future.

\textit{Acknowledgement.}\textemdash  The authors are grateful to Xiao-Rui Lyu and Hai-Long Ma for the measurement of the ratio of up-down asymmetries  at BESIII, to Ji-Bo He,  Liang Sun, Wen-Bin Qian and Yue-Hong Xie for the LHCb measurement,  and to Emi Kou for useful discussions. In particular we thank Prof.~Yue-Hong Xie for discussing the separation of $K_1(1270)$ contributions in experiment.  This work is supported in part by Natural Science Foundation of China under
grant No. 11575110, 11735010, 11911530088, 11975112, U1732101,  by Natural Science Foundation of Shanghai under grant No. 15DZ2272100, and by Gansu Natural Science Fund under grant No.18JR3RA265. 


\end{document}